\documentclass{elsart}
\usepackage{graphicx}
\newcommand{\ttbar}{$t\bar{t} \;$}
\newcommand{\bbbar}{$B/\bar{B} \;$}
\newcommand{\jpsi}{J$/\psi \;$}
\newcommand{\lj}{$\ell$J$/\psi \;$}
\newcommand{\mmax}{M$^{max} \;$}
\newcommand{\mmaxl}{M$^{max}_{\ell\mathrm{J}/\psi} \;$}
\newcommand{\pyth}{P{\footnotesize YTHIA} }
\newcommand{\herw}{{\small HERWIG} }
\def\gsim{\mathrel{\rlap{\raise.4ex\hbox{$>$}} {\lower.6ex\hbox{$\sim$}}}}
\def\lsim{\mathrel{\rlap{\raise.4ex\hbox{$<$}} {\lower.6ex\hbox{$\sim$}}}}

\begin{document}

\begin{frontmatter}
\title{Top mass determination in leptonic final states with J$/\psi$}
\author{Avto Kharchilava}
\address{Institute of Physics, Georgian Academy of Sciences,
Tbilisi, Georgia}

\begin{abstract}
Taking advantage of large top production rates at the LHC,
the leptonic final states with J$/\psi$ are explored for a precise
determination of the top quark mass. The top is partially reconstructed by
combining the isolated lepton from the W with the J$/\psi$ from the decays
of the corresponding $b$-quark. The method relies, to a large extent, on
the proper Monte-Carlo description of top production and decay.
The main emphasis is put on the expected systematics uncertainties.
Mass measurement accuracy is dominated by the current understanding of
theoretical uncertainties which result in a systematic error of
$\lsim$~1~GeV.

\medskip

\hspace*{-4mm}{\it PACS:} 14.65.Ha, 12.38.-t;
{\it keywords:} Top quark mass, LHC
\end{abstract}
\end{frontmatter}

\section{Introduction}

One of the main interests in the top physics studies
at the LHC will be an accurate measurement of the top mass. Being
a fundamental parameter of the Standard Model (SM) it constraints, along
with the W, the Higgs mass thus providing an excellent consistency
check of the SM.

Currently the best Tevatron single experiment results on the top mass
are obtained with the lepton plus jets final states which yield:
M$_{top}$ = 175.9 $\pm$ 4.8 (stat.) $\pm$ 5.3 (syst.) (CDF)~\cite{cdf}
and 173.3 $\pm$ 5.6 (stat.) $\pm$ 5.5 (syst.) (D\O)~\cite{d0}.
The systematic errors in both measurements are largely dominated by the
uncertainty on the jet energy scale and amount to 4.4~GeV and 4~GeV
for the CDF and D\O , respectively. On the other hand, the systematics
in di-lepton channels are somewhat less, but the statistical errors
are significantly larger, by a factor of $\gsim$~2, as compared to the
lepton plus jets final states.
In future runs of the Tevatron with about 20-fold increase in statistics
the top mass will be measured with an accuracy of
$\sim$~2~GeV~\cite{future}; in the lepton
plus jets channel the error is dominated by the systematics
while in di-lepton channels the limiting factor would still be statistics.

\begin{figure}[hbtp]
\vspace*{0.3cm}
  \begin{center}
    \resizebox{8cm}{!}{\includegraphics{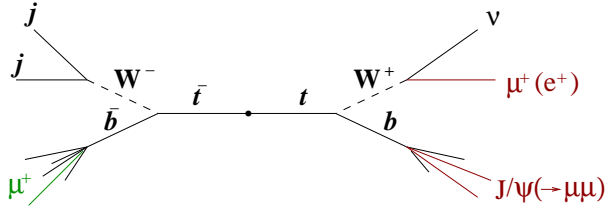}}
\caption{Schematic of the top decay to leptonic final states with
J$/\psi$.}
    \label{topc2}
  \end{center}
\end{figure}

The legitimate question to ask -- what is the ultimate precision
on the top mass that can be attained at the LHC? Apparently, to
significantly improve on jet measurements at LHC would be a challenge.
However, one can take the advantage of large top production rates and
explore leptonic final states.
In this paper we consider four- (three-) lepton final states involving
the \jpsi$\to \mu^+\mu^-$ decays, see Fig.~\ref{topc2}.

Here, the top mass is partially reconstructed by combining the isolated
lepton, $\mu$ or e, from the W decays with the \jpsi from the decays of
the corresponding $b$-quark, i.~e. reconstructing the lepton
plus \jpsi invariant mass, M$_{\ell\mathrm{J}/\psi}$~\cite{LoI,note99}.
In order to determine the top decay topology one can require an additional
tagging muon-in-jet coming from other $b$. The branching suppression
factor is $5.3\times10^{-5}$ taking into account the charge conjugate
reaction and W~$\to$~e$\nu$ decays. In spite of rather low branching
ratios, we stress that these final states are experimentally very clean
and can be exploited even at highest LHC luminosities. Furthermore,
one can also explore an other way to associated the \jpsi with the
corresponding isolated lepton -- by measuring the jet charge of identified
$b$('s) and not requiring the tagging muon.
In this case we gain a factor of more than 10 in statistics as compared to
the four-lepton final states.

Similar approaches, namely the use of the correlation between the
top mass and its decay products, e.~g. in final states with an isolated
lepton combined with the muon from the semileptonic decays of the
corresponding $b$'s, have been explored in earlier studies
\cite{ssc,aachen}. One should say that all these methods of
top mass determination essentially rely on the Monte-Carlo description of
its production and decay. Nonetheless, the model, to a great extent,
can be verified and tuned to the data.

\section{Analysis}

In the following we assume the \ttbar production cross-section of 800~pb
for M$_{top}$ = 175~GeV.
Events are simulated with the \pyth 5.7 \cite{pyth} or
\herw 5.9~\cite{herw} event generators. Particle momenta are
smeared according to parameterizations obtained from detailed
simulation of the CMS detector performance. Four-lepton events are
selected by requiring an isolated lepton of $p_T>$ 15~GeV in central
pseudorapidity range of $|\eta|<$ 2.4 and three non-isolated, centrally
produced muons of $p_T>$ 4~GeV in $|\eta|<$ 2.4, with the invariant mass
of the two of them being consistent with the \jpsi mass. These cuts would
significantly reduce the external (non-\ttbar) background, mainly the
W$b\bar{b}$ production, \footnote{No reliable event generator
is currently available to simulate the W$b\bar{b}$ process. \pyth
results indicate that with the above cuts this source of the background
can be kept at a per cent level.}
which can be further reduced by employing, in addition, two central jets
from another W, if needed.
Resulting kinematical acceptance of the selection criteria is 0.3; this
rather small value is largely due to soft muons from J$/\psi$ and $b$.
In one year high luminosity running of LHC, corresponding to an $L_{int} =
10^5$~pb$^{-1}$, and assuming the trigger plus reconstruction efficiency
of 0.8, we expect about $10^5 \times 800 \times 5.3\cdot10^{-5} \times 0.3
\times 0.8 = 1000$ events.

\begin{figure}[hbtp]
\vspace*{-1.2cm}
\begin{minipage}{.48\linewidth}
  \begin{center}
    \resizebox{7.4cm}{!}{\includegraphics{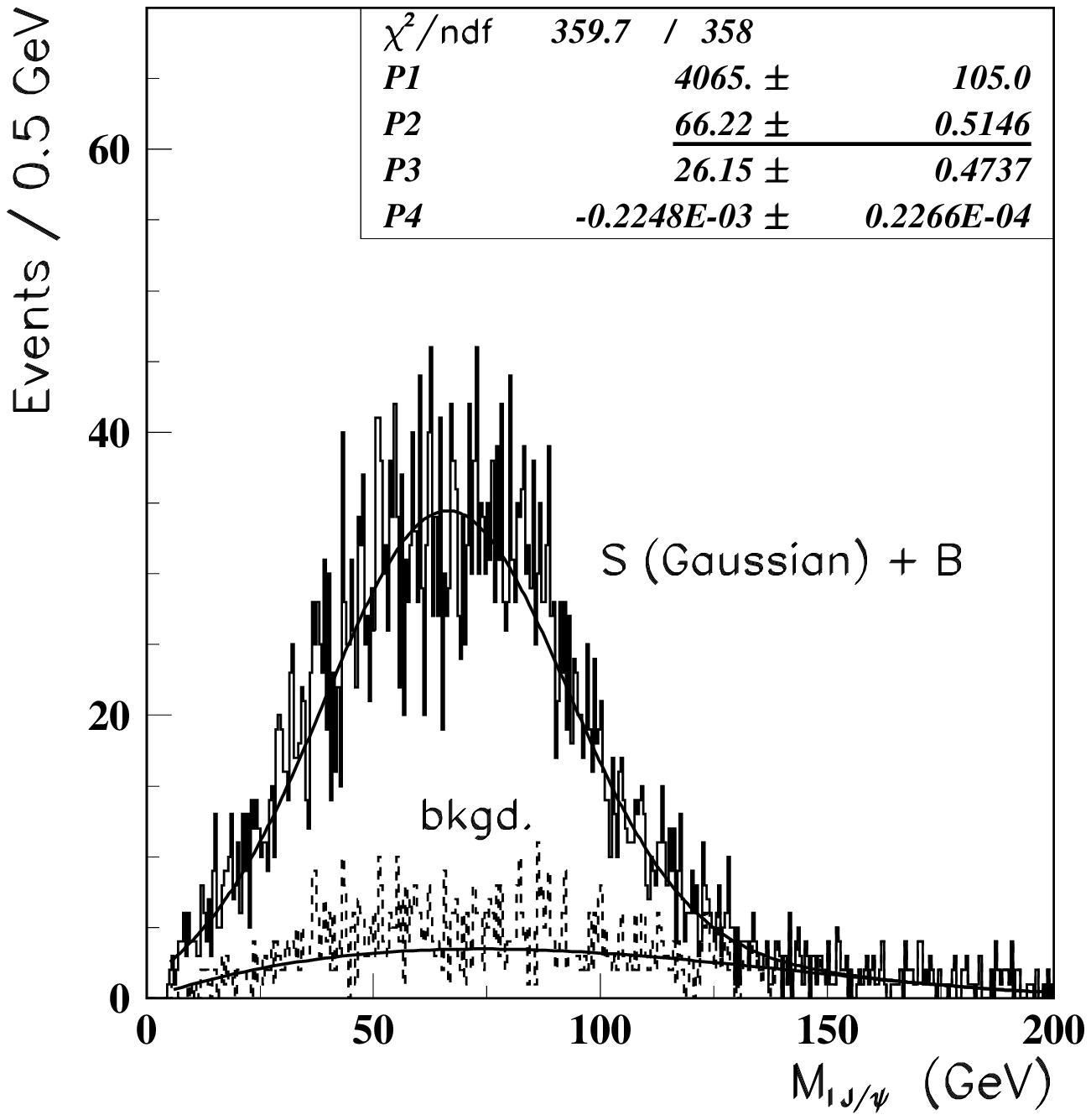}}
\vspace*{-1.2cm}
\caption{Example of the \lj invariant mass spectrum in 4-lepton
final states. Number of events correspond to four years running at LHC
high luminosity.}
\label{sbac}
  \end{center}
\end{minipage}
\hfill
\begin{minipage}{.48\linewidth}
  \begin{center}
    \resizebox{7.4cm}{!}{\includegraphics{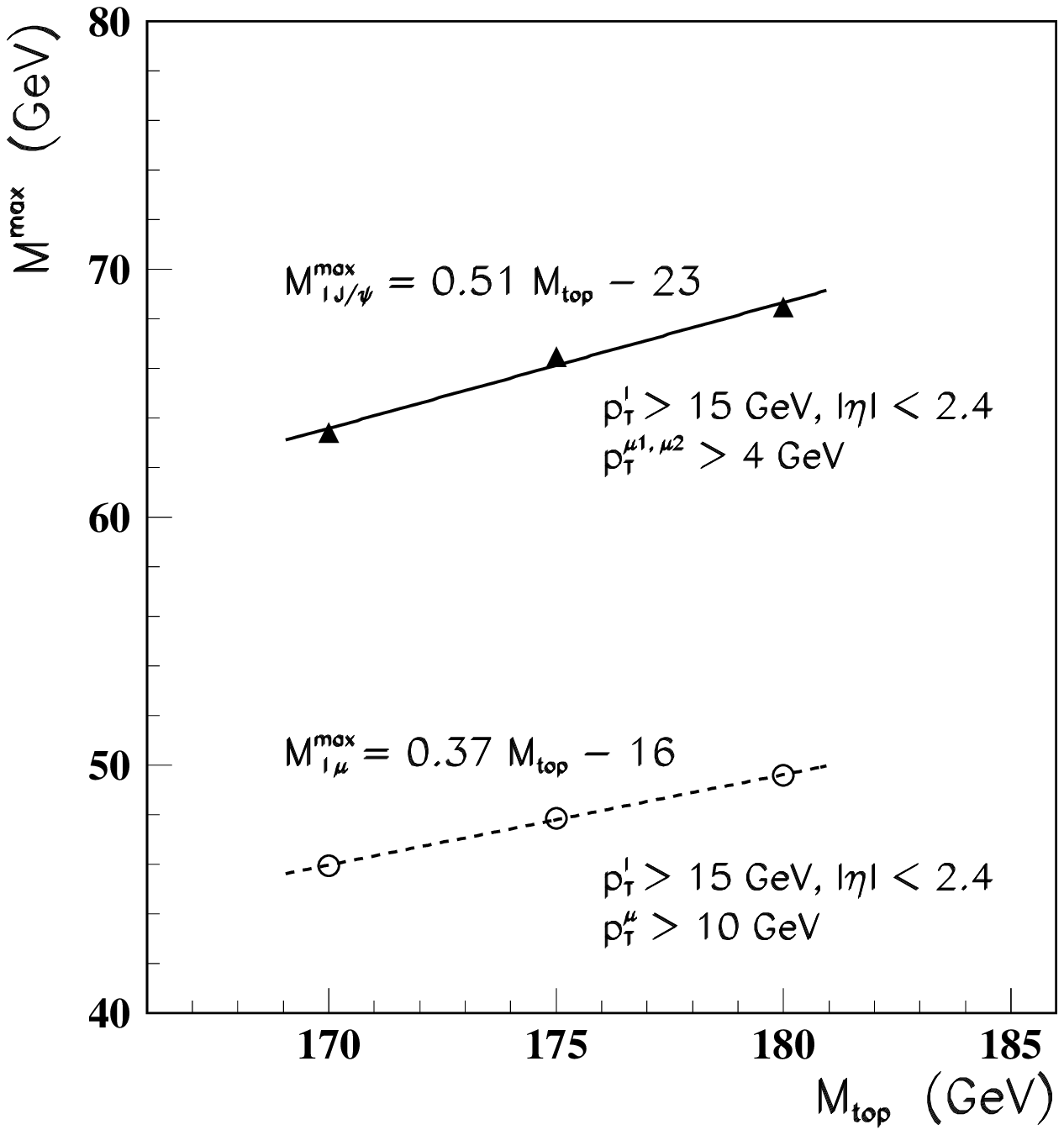}}
\vspace*{-1.2cm}
  \caption{Correlation between \mmax and the top-quark mass in isolated
lepton plus \jpsi (solid line) and isolated lepton plus $\mu$-in-jet
(dashed line) final states.}
\label{fitl2}
  \end{center}
\end{minipage}
\end{figure}

An example of the \lj mass distribution with the expected background
is shown in Fig.~\ref{sbac}. The background is internal (from the \ttbar
production) and is due to the wrong assignment of the \jpsi to the
corresponding isolated lepton. These tagging muons of wrong sign
are predominantly originating from \bbbar oscillations,
$b \to c \to \mu$ transitions,
W$(\to c, \ \tau) \to \mu$ decays, $\pi/$K decays in flight and amount to
$\sim$~30\% of the signal combinations. The shape of the signal \lj
events (those with the correct sign of the tagging muon) is consistent
with a Gaussian distribution over the entire mass interval up to its
kinematical limit of $\sim$~175 GeV. The background shape is approximated
by a polynomial function of degree~3. The parameters of this polynomial
function are determined with ``data'' made of the wrong
combinations of \lj with an admixture of signal. In such a way
the shape of the background is determined more precisely and in situ.
Thus, when the signal distribution is fitted, only the background
normalization factor is left as a free parameter along with the three
parameters of a Gaussian. The result of the fit
is shown in Fig.~\ref{sbac}. We point out that this procedure allows to
absorb also the remaining external background (if any) into the background
fit function.

As a measure of the top quark mass we use the mean value
(position of the maximum of the distribution) of the Gaussian, \mmaxl.
In four years running at LHC with high luminosity the typical errors on
this variable, including the uncertainty on the background, are about
0.5~GeV. It is composed of $\lsim$~0.5~GeV statistical error and
$\lsim$~0.15~GeV systematics contribution due to the uncertainty on the
measurement of the background shape.

The measurement of the \mmaxl can then be related to the generated top
quark mass.
An example of the correlation between the M$^{max}$ and M$_{top}$ is
shown in Fig.~\ref{fitl2} along with the parameters of a linear fit.
For comparison, we also show the corresponding dependence in a more
traditional isolated lepton plus $\mu$-in-jet channel.
Not surprisingly, the stronger correlation, and thus a better sensitivity
to the top mass, is expected in the \lj final states
as compared to the isolated lepton plus $\mu$-in-jet channel. This is
because, in the former case, we pickup a heavy object (i.~e. \jpsi)
from the $b$-jet which carries a larger fraction of its momentum.
Apparently, when determining the top mass, the \mmaxl measurement error,
statistical and systematic, would scale up as the inverse slope value
of the fit which is a factor of 2 in our case. Hence the statistical error
on the top mass in this particular example would be $\sim$~1~GeV.

It is appropriate to comment on the ways to obtain a larger event sample.
Possibilities might include reconstruction of the J$/\psi \to$~e$^+$e$^-$
decays in $b$-jets \cite{jpsiee} and/or relaxing
the lepton $p_T$ thresholds \cite{note99}.
An even larger event sample can be obtained in three lepton final states,
i.~e. without the tagging muon, but using instead the jet charge technique
to determine the \ttbar decay topology. The jet charge is defined as a
$p_T$-weighted charge of particles collected in a cone around the \jpsi
direction. Obviously, this kind of analysis requires detailed simulations
with full pattern recognition.
However, particle level simulations performed with \pyth and with
realistic assumptions on track reconstruction efficiency indicate
comparable to muon-tag performance figures, but with about 10 times less
integrated luminosity.
In any case, through the LHC lifetime, one can collect enough
events so that the overall top mass measurement accuracy would not
be hampered by the lack of statistics; it would rather be limited
by the systematics uncertainties which are tightly linked with the
Monte-Carlo tools in use, as will be argued in the following section.

\section{Systematics}

Let us first briefly mention the features of \pyth and \herw
that are relevant to our analysis. They both incorporate only the
leading order matrix elements for \ttbar production and decay,
and simulate the QCD final state effects via parton showers. This
algorithm is not exact at higher orders in $\alpha_s$ but could be, in
principle, matched to the next-to-leading order calculations.
Both event generators keep track of color connections, however the
hadronization is performed in different ways: in \pyth ({\small JETSET})
it is done via a string fragmentation model while \herw exploits a
cluster hadronization scheme. Eventually, free parameters of these event
generators are tuned to reproduce the data, especially from e$^+$e$^-$
colliders. These tunings to data usually include event shape variables,
identified particles production rates, their momentum spectra, etc.

An essential aspect of the current analysis is to understand
limitations which would arise from the Monte-Carlo description of the
top production and decay. With \pyth we have investigated the following
sources of systematic uncertainties on the \mmaxl measurements:

\begin{enumerate}
\item{\it Initial State Radiation.} \\
The \mmaxl value is unchanged even switching the initial state radiation
off.

\item{\it Final State Radiation (FSR).} \\
Large shift of $\sim$~7~GeV is observed when it is switched off.
To evaluate the uncertainty we varied the parton virtuality scale
$m_{min}$ -- an invariant mass cut-off below which the showering is
terminated. A $\pm$50\% variation of it around the default (tuned to data)
value of 1~GeV induces an uncertainty of $^{+0.1}_{-0.15}$~GeV.

\item
{\it QED bremsstrahlung.} \\
Only a small effect is observed when it is switched off.
Furthermore, QED radiation is well understood and can be
properly simulated.

\item
{\it Parton distribution functions (PDF).} \\
Depending on the relative fraction of gluon/quarks versus $x$
in various PDFs the top production kinematics might be different.
No straightforward procedure is available for the moment
to evaluate uncertainties due to a particular choice of PDF.
Here we compare results obtained with the default set {\small CTEQ2L}
\cite{cteq2l}
and a more recent {\small CTEQ4L} \cite{cteq4l} parameterizations of PDFs.
The observed change in the \mmaxl value is well within 0.1~GeV.

\item
{\it Top $p_T$ spectrum.} \\
One does not expect significant uncertainties associated with the
top $p_T$ spectrum, i.~e. its generated shape. This is primarily due to
a good agreement between the parton shower Monte-Carlos and the
next-to-leading order QCD calculations over a wide range of $p_T$ up to
1~TeV. In Ref.~\cite{mlm1} this comparison is done with the \herw event
generator. Similar results we have obtained also with P{\footnotesize
YTHIA}. However, to see an effect we have artificially altered the top
$p_T$ spectrum by applying a cut at the generator level. For example, a
1$\sigma$ effect shows up only when the $\hat{p}_T$ cut-off pushed up to
100~GeV.

\item
{\it Top and W widths.} \\
Kinematical cuts that one applies usually affect the observed Breit-Wigner
shape (tails) of decaying particles. Conversely, poor knowledge of the
widths may alter the generated \lj mass spectrum depending on the cuts.
Only little change in the \mmaxl value is seen even with the zero-width
approximation.

\item
{\it W polarization.} \\
A significant shift is found for the isotropic decays of W
when compared to the SM expectation of its $\sim 70$\% longitudinal
polarization. In future runs of the Tevatron the W polarization
will be measured with a $\sim 2$\% accuracy \cite{future}, and at the LHC
this would be further improved so that it should not introduce
additional uncertainties in simulations.

\item
{\it \ttbar spin correlations.} \\
A ``cross-talk'' between $t$ and $\bar{t}$ decay products is
possible due to experimental cuts. To examine this effect in details
the $2 \to 6$ matrix elements have been implemented in \pyth
preserving the spin correlations~\cite{serg}. No sizeable difference
in the \mmaxl value is seen compared to the default $2 \to 2$ matrix
elements.

\item
{\it $b$ fragmentation, except FSR.} \\
As a default, in \pyth we have used the Peterson form for the $b$-quark
fragmentation function with $\varepsilon_{b}=0.005$.
Variation of this value by $\pm$10\%~\cite{lephfwg} leads to an
uncertainty of $^{-0.3}_{+0.25}$ GeV.

\item
{\it Background.} \\
The uncertainty would be mainly due to an inaccurate measurement
of the background shape and the systematics contribution of
$\lsim$~0.15~GeV quoted in previous section would scale down
with increasing statistics. For example, already with $\sim 10^4$
events the induced uncertainty is $\lsim$~0.1~GeV.

\item
{\it Detector resolution.} \\
Here we have considered only Gaussian smearing of particle momenta
and the effect on the M$^{max}_{\ell\mathrm{J}/\psi}$ measurement
uncertainty is negligible. A possible nonlinearity of the detector
response can be well controlled having a huge sample of J$/\psi$, Y and Z
leptonic decays that will be available at LHC.

\end{enumerate}

\begin{figure}[hbtp]
\vspace*{-1.6cm}
  \begin{center}
    \resizebox{10cm}{!}{\includegraphics{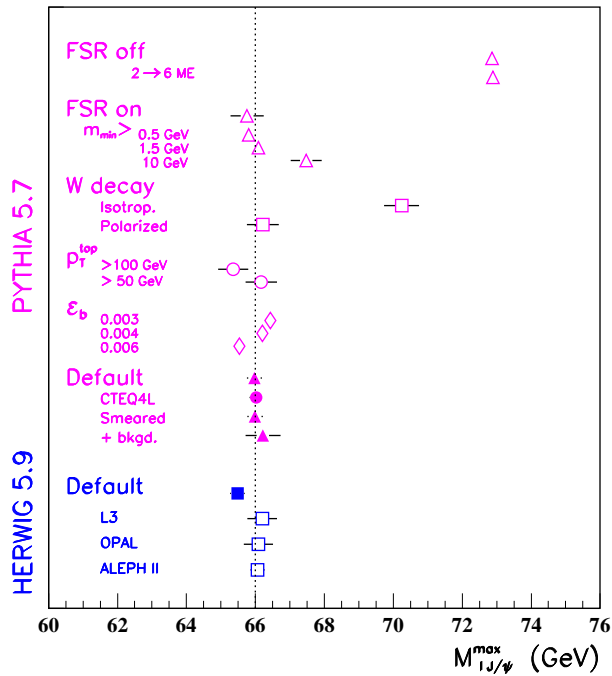}}
\vspace*{-1.2cm}
    \caption{Summary on the systematics studies.}
    \label{syst}
  \end{center}
\end{figure}

A summary of these studies is given in Fig.~\ref{syst}.
One sees an impressive stability of results for reasonable choices
of parameters. The expected systematic error in the \mmaxl determination
is $\lsim$~$^{+0.3}_{-0.4}$~GeV which translates into a systematic
error on the top mass of $\delta$M$_{top}\lsim$~$^{+0.6}_{-0.8}$~GeV.

\medskip

Another way to estimate systematics uncertainties could be by a comparison
of \pyth and \herw expectations. With \herw we have tried various tunings
from LEP experiments as well as its default settings listed in
Table~\ref{hwset}. They all yield comparable results to each other and to
\pyth results, and are within $\lsim$~0.5~GeV. \footnote{This should
be taken with a precaution -- bugs have been found in the treatment
of the top decays in \herw~\cite{corc} -- so its impact on current
analysis should be evaluated once the improved version of \herw is
publicly available.}
This corresponds to a systematic uncertainty
$\delta$M$_{top}~\lsim$~1~GeV.

\vspace*{3mm}
\begin{table}[ht]
\centering
\caption{\herw parameter settings \cite{herw}.}
\label{hwset}
\vspace*{2mm}
\begin{tabular}{|l|c|c|c|c|}
\hline
Parameter & Default & {\small ALEPH II} & {\small L3} & {\small OPAL} \\
\hline
\hline
$\Lambda_{QCD}$ & 0.18    & 0.173    & 0.177 & 0.15 \\
$m_{g}^{eff.}$  & 0.75    & 0.645    &       & 0.7  \\
{\small CLMAX}  & 3.35    & 3.025    & 3.006 & 3.75 \\
{\small CLPOW}  & 2.      &          & 2.033 & 1.3  \\
{\small PSPLT}  & 1.      & 0.984    &       & 0.85 \\
{\small CLSMR}  & 0.      &          & 0.35  &      \\
\hline
\end{tabular}
\end{table}
\vspace*{3mm}

One more feature specific to top decays is related to the matrix element
corrections to parton shower simulations~\cite{corc}.
A significant shift is observed in the mean value of the isolated lepton
plus $B$-hadron invariant mass spectra when these corrections are
implemented in {\small HERWIG}~\cite{mlm2}. On the other hand, this shift
does not necessarily induces an additional systematic error (which is at
least partially accounted for in 2 -- the FSR effects) once it is properly
taken into account. Further studies are required when the corresponding
implementation is publicly made available.

\section{Conclusions}

Abundant production of top quarks at LHC allows to use leptonic final
states to precisely determine the top quark mass thus avoiding limitations
that are ultimately associated to measurements with jets.

\begin{enumerate}
\item[$\bullet$]
In $\ell$J$/\psi$ final states the top mass can be determined
with a systematic uncertainty of $\lsim$~1~GeV.

\item[$\bullet$]
These final states are experimentally very clean and can be exploited even
at highest LHC luminosities.

\item[$\bullet$]
The method heavily relies on the Monte-Carlo description of top production
and decay, however the model can be verified and tuned to the data.

\item[$\bullet$]
The precision would be limited by the theoretical uncertainties
which is basically reduced to the one associated with the $t \to b$
transition.

\end{enumerate}

To conclude, this method of top mass determination looks very promising
and particularly well suited for the CMS detector with its 4~Tesla field
and ensuring precise lepton measurements.

\vspace{1cm}
{\bf Acknowledgements}
\vspace{3mm}

This work has been carried out in the course of the
Workshop on Standard Model Physics (and more) at the LHC.
I am thankful to Daniel Denegri, Michelangelo Mangano and John Womersley
for their constant interest to this analysis and encouragement.


\begin{thebibliography}{9}
\bibitem{cdf} CDF Collaboration, Phys. Rev. Lett. 80 (1998) 2767;
Phys. Rev. Lett. 82 (1999) 271.
\bibitem{d0} D\O\ Collaboration, Phys. Rev. D58 (1998) 052001.
\bibitem{future} R.~Frey {\it et al.}, FERMILAB-CONF-97/085.

\bibitem{LoI} CMS Letter of Intent, CERN/LHCC 92-3 (1992) 90; \\
I.~Iashvili {\it et al.}, CMS TN/92-34 (1992).

\bibitem{note99} A.~Kharchilava, CMS Note 1999/065 (1999).

\bibitem{ssc} K.~J.~Foley {\it et al.}, in Proc. of the Workshop on
Experiments, Detectors, and Experimental Areas for the Supercollider,
Berkeley, 1987, eds. R.~Donaldson and M.~G.~D.~Gilchriese
(World Scientific, 1988) 701.

\bibitem{aachen} G.~Unal and L.~Fayard, in Proc. of Large Hadron
Collider Workshop, Aachen, 1990, eds. G.~Jarlskog and D.~Rein, CERN 90-10,
vol.~II (1990) 360.

\bibitem{pyth} T.~Sj\"{o}strand, Comp. Phys. Comm. 82 (1994) 74.

\bibitem{herw} G.~Marchesini {\it et al.}, Comp. Phys. Comm. 67
(1992) 465; \\
Current public version 5.9 of \herw and various parameter tunings are
taken from this address:
http://hepwww.rl.ac.uk/theory/seymour/herwig/.

\bibitem{jpsiee} A.~Kharchilava and P.~Pralavorio, CMS TN/96-116 (1996).

\bibitem{cteq2l}
H.~L.~Lai {\it et al.}, Phys. Rev. D51 (1995) 4763.

\bibitem{cteq4l}
H.~L.~Lai {\it et al.}, Phys. Rev. D55 (1997) 1280.

\bibitem{mlm1}
M.~Mangano, talk given at the Workshop on Standard Model Physics
(and more) at the LHC, CERN, May 25, 1999,
http://home.cern.ch/$\sim$mlm/lhc99/ lhcworkshop.html; \\
S.~Frixione, M.~Mangano, P.~Nason and G.~Ridolfi,
Phys.~Lett. B351 (1995) 555.

\bibitem{serg}
S.~R.~Slabospitsky, in preparation; implemented into \pyth
in collaboration with L.~Sonnenschein.

\bibitem{lephfwg}
LEP Heavy Flavour Working Group, LEPHF/98-01; the $\pm$10\%
uncertainty on $\varepsilon_{b}$ is inferred from LEP/SLD
precision of $\sim 1$\% on average scaled energy of $B$-hadrons.

\bibitem{corc} G.~Corcella and M.~H.~Seymour, Phys. Lett. B442 (1998) 417.

\bibitem{mlm2}
M.~Mangano, talk given at the Workshop on Standard Model
Physics (and more) at the LHC, CERN, October 25, 1999,
http://home.cern.ch/$\sim$mlm/lhc99/ lhcworkshop.html; \\
G.~Corcella, hep-ph/9911477; \\
G.~Corcella, M.~Mangano and M.~H.~Seymour, in preparation.

\end{thebibliography}
\end{document}